\documentclass[aps, prb, twocolumn, showpacs, amsmath, amssymb]{revtex4}
\usepackage{graphicx}
\usepackage{bm}

\begin{document}

\title{Local defect in a magnet with long-range interactions}

\author{Jos\'{e} A. Hoyos}
\email{hoyosj@umr.edu} \affiliation{Department of Physics, University of Missouri-Rolla,
Rolla, Missouri, 65409, USA.}

\author{Thomas Vojta}
\email{vojtat@umr.edu} \affiliation{Department of Physics, University of Missouri-Rolla,
Rolla, Missouri, 65409, USA.}

\date{\today}
\pacs{75.10.Jm, 75.10.Nr, 75.40.-s}

\begin{abstract}
We investigate a single defect coupling to the square of the order parameter in a nearly
critical magnet with long-range spatial interactions of the form $r^{-(d+\sigma)}$,
focusing on magnetic droplets nucleated at the defect while the bulk system is in the
paramagnetic phase. To determine the static droplet profile, we solve a
Landau-Ginzburg-Wilson action in saddle point approximation. Because of the long-range
interaction, the droplet develops a power-law tail which is energetically unfavorable.
However, as long as $\sigma>0$, the tail contribution to the droplet free energy is
subleading in the limit of large droplets; and the free energy becomes identical to the
case of short-range interactions. We also consider the effects of fluctuations and find
that they do not change the functional form of the droplet as long as the bulk system is
noncritical. Finally, we study the droplet quantum dynamics with and without dissipation;
and we discuss the consequences of our results for defects in itinerant quantum
ferromagnets.
\end{abstract}
\maketitle

\section{Introduction}

In a nearly critical system, a local defect that prefers the ordered phase can induce the
nucleation of a droplet of local order in the nonordered background. Such droplets arise,
e.g., in disordered systems due to the presence of rare strongly coupled spatial regions.
They can have surprisingly strong consequences for the properties of the phase transition.
In a classical magnet at nonzero temperature, a large finite-size droplet does not have a
static magnetization; instead it fluctuates very slowly because flipping the droplet
requires coherently changing the order parameter in a large volume. More than 30 years
ago, Griffiths \cite{Griffiths69} showed that rare regions and the magnetic droplets
formed on them, lead to a singularity in the free energy in a whole temperature region
above the critical point, which is now known as the Griffiths region or the Griffiths
phase.\cite{RanderiaSethnaPalmer85} Later, it was shown that this singularity is only an
essential one \cite{Wortis74,Harris75,BrayHuifang89} and thus probably unobservable in
experiment (see also Ref.\ \onlinecite{Imry77}).

The effects of magnetic droplets are greatly enhanced if the underlying defects are
extended macroscopic objects (linear or planar defects). In these cases, the droplet
dynamics is even slower and so increases their effects. This was first found in the
McCoy-Wu model, \cite{McCoyWu68,McCoyWu68a} a 2D Ising model with linear defects. Later
it was studied in great detail in the context of the quantum phase transition of the
random transverse-field Ising model where the defects are extended in the imaginary time
dimension. \cite{Fisher92,Fisher95} In these systems, the Griffiths singularity in the
free energy actually takes a power-law form, and the susceptibility diverges inside the
Griffiths region.

Recently, it has been shown that even stronger effects than these power-law Griffiths
singularities can occur in Ising magnets with planar
defects.\cite{Vojta03b,SknepnekVojta04} Droplets that are extended in two dimensions can
undergo the magnetic phase transition (and develop a static order parameter)
independently from the bulk system. This leads to a destruction of the global sharp phase
transition by smearing. (Note that an unusual magnetization-temperature relation was
already found in the numerical mean-field analysis, Ref.\ \onlinecite{BBIP98}, but it was
interpreted as power-law critical behavior with a very large exponent $\beta$.) Similar
smeared phase transitions have also been found in a non-equilibrium system in the
presence of linear defects \cite{Vojta04}. A recent review of these and other rare region
effects can be found in Ref.\ \onlinecite{Vojta06}.

One particularly interesting class of problems concerns droplets in metallic quantum
magnets. In these systems, the dynamics of the magnetic modes is overdamped because they
couple to gapless fermionic excitations. In metallic (Ising) antiferromagnets, this
dissipative environment strongly suppresses tunneling, and sufficiently large droplets
completely freeze at low temperatures.
\cite{CastroNetoJones00,MillisMorrSchmalian01,MillisMorrSchmalian02} The global quantum
phase transition is thus smeared. \cite{Vojta03a}

In metallic ferromagnets, the situation is further complicated because the coupling
between the magnetic modes and the gapless fermionic degrees of freedom generates an
effective long-range spatial interaction between the magnetic fluctuations.
\cite{VBNK96,VBNK97,BelitzKirkpatrickVojta97} This interaction which takes the form
$r^{-(2d-1)}$ for clean electrons and $r^{-(2d-2)}$ for diffusive electrons, where $d\ge
2$ is the spatial dimensionality, can be viewed as a result of generic scale invariance
(for a recent review see Ref.\ \onlinecite{BelitzKirkpatrickVojta05}). Understanding
defects in nearly critical metallic quantum ferromagnets thus leads to the question of
whether the existence and the properties of magnetic droplets are influenced by the
long-range spatial interaction.

In this paper, we therefore develop the theory of a single defect coupling to the square
of the order parameter in a nearly critical classical or quantum magnet with power-law
spatial interactions of the form $r^{-(d+\sigma)}$ with $\sigma > 0$ to ensure a proper
thermodynamic limit. A crucial effect of the long-range interactions is that the tail of
the magnetic droplet decays into the bulk region like a power-law of the distance as
mandated by Griffiths theorem. \cite{Griffiths67} Such a strong tail extending into the
region that prefers the disordered phase can be expected to be energetically unfavorable.
To find out to what extent this hinders the formation of the magnetic droplet, we study
the droplet free energy within the saddle-point approach. In the quantum case, we also
consider the tunneling dynamics of the droplet for three cases: undamped dynamics,
overdamped dynamics due to Ohmic dissipation, and a conserved overdamped dynamics as in
the itinerant ferromagnet.

Our paper is organized as follows: We introduce our model, a classical or quantum
$\phi^4$-theory with long-range spatial interactions, in Sec.~\ref{sec:The-Model}. In
Sec.~\ref{sec:Droplet-static-profile}, we analyse the free energy of a droplet within
saddle-point approximation, and we discuss fluctuations. The droplet dynamics in the
quantum case is considered in Sec.~\ref{sec:The-dynamics}. The concluding
Sec.~\ref{sec:conclusions} is devoted to a summary as well as a discussion of the order
parameter symmetry and the consequences of our results for quantum Griffiths effects.

\section{The Model\label{sec:The-Model}}

In this section we introduce our model, a $d$-dimensional Landau-Ginzburg-Wilson field
theory with long-range power-law interactions for a scalar order parameter field
$\varphi$. We first formulate the model for the case of a zero-temperature quantum phase
transition, and we later discuss the necessary changes for a classical thermal phase
transition. The action of our quantum $\phi^4$-theory reads
\begin{equation}
S=S_{\rm stat}+S_{\rm dyn},\label{eq:total_action}
\end{equation}
with the static part given by
\begin{eqnarray}
S_{\rm stat} &=& \int d\tau \int d\mathbf{x}
d\mathbf{y}\varphi\left(\mathbf{x},\tau\right)
             \Gamma\left(\mathbf{x},\mathbf{y}\right)\varphi\left(\mathbf{y},\tau\right)
             \nonumber\\
         &+&\frac{u}{2}\int d\tau d\mathbf{x}\varphi^{4}\left(\mathbf{x},\tau\right)~.
\label{eq:S_stat}
\end{eqnarray}
Here, $\mathbf{x}$ and $\mathbf{y}$ are position vectors and $\tau$ is imaginary time.
The bare two-point vertex, $\Gamma\left(\mathbf{x},\mathbf{y}\right)= \Gamma_{\rm
NI}(\mathbf{x})\delta\left(\mathbf{x}-\mathbf{y}\right) +\Gamma_{\rm
I}\left(\mathbf{x},\mathbf{y}\right)$, contains a non-interacting part and the attractive
long-range interaction. The latter is given by
\begin{equation}
\Gamma_{\rm I}\left(\mathbf{x},\mathbf{y}\right)=
     -\gamma\left[\xi_{0}^{2}+\left|\mathbf{x}-\mathbf{y}\right|^{2}\right]^{-\left(\frac{d+\sigma}{2}\right)}.
     \label{eq:I_kernel}
\end{equation}
Here, $\gamma$ is the interaction strength, $\xi_0$ is a microscopic cutoff length scale
of the order of the lattice constant, and $\sigma$ controls the range of the interaction.
To ensure a proper thermodynamic limit (an extensive free energy), $\sigma$ must be
positive. Note that an additional short-range interaction of the usual form
$|\nabla\varphi|^2$ can be added, if desired. As will be shown in Sec.\
\ref{subsec:SP-equation}, its contribution is subleading. The noninteracting part of the
vertex reads
\begin{equation}
\Gamma_{\rm NI}\left(\mathbf{x}\right)=t_{0}+\delta t\left(\mathbf{x}\right)+\Gamma_{0},
\label{eq:NI_kernel}
\end{equation}
where $t_{0}$ is the bulk distance from criticality,\footnote{In principle, one must
distinguish between the bare and the renormalized distance from the critical point. We
will suppress this difference unless otherwise noted, because it is of no importance for
our considerations.}
 and the constant $\Gamma_{0}$ is chosen
such that it cancels the $(\mathbf{q}=0)$ Fourier component of the interaction (thus ensuring
that the bulk critical point is indeed at $t_{0}=0$). It takes the value
$\Gamma_0=\Omega_d\gamma\xi_0^{-\sigma}\,B(\sigma/2,d/2)/2$.
Here $\Omega_d$ is the surface of a $d$-dimensional unit sphere, and $B(x,y)$ is Euler's beta function.
$\delta t\left(\mathbf{x}\right)$
is the defect potential. For definiteness we consider a single spherically symmetric defect at
the origin,
\begin{equation}
\delta t (\mathbf{x}) = \left\{\begin{array}{rr} -V & \quad (|\mathbf{x}|<a) \\ 0 & (|\mathbf{x}|>a) \end{array}
\right.~.
\label{eq:defect}
\end{equation}
We are interested in the case $V>0$, i.e., in defects that favor the ordered phase.

When discussing the quantum tunneling dynamics of the magnetic droplets in
Sec.~\ref{sec:The-dynamics}, we will compare three different dynamical actions.
(i) In the undamped case, the dynamical action is given by
\begin{equation}
S_{\rm dyn}^{(1)}=  T\sum_{\omega_{n}}\int d\mathbf{q}~\frac {\omega_{n}^2} {c^2}
\left|\tilde{\varphi}\left(\mathbf{q},\omega_{n}\right)\right|^{2} \label{eq:S_dyn_z1}
\end{equation}
where $\tilde{\varphi}(\mathbf{q},\omega_{n})$ is the Fourier transform of the order
parameter field in terms of wave number $\mathbf{q}$ and Matsubara frequency $\omega_n$,
$T$ is the temperature, and $c$ plays the role of a velocity of the undamped modes.

(ii) If the magnetic modes are coupled to an ohmic bath, the leading term in the dynamic action
takes the form
\begin{equation}
S_{\rm dyn}^{(2)}=\tilde\alpha T\sum_{\omega_{n}}\int d\mathbf{q}~\left|\omega_{n}\right|
\left|\tilde{\varphi}\left(\mathbf{q},\omega_{n}\right)\right|^{2}, \label{eq:S_dyn_z2}
\end{equation}
where $\tilde\alpha$ measures the strength of the dissipation, and there is a microscopic
frequency cutoff $|\omega_n| < \omega_{\rm mic}$.

(iii) Finally, we also consider the case of overdamped dynamics with order parameter conservation
analogous to the itinerant ferromagnet. The leading term in the dynamic action is given by
\begin{equation}
S_{\rm dyn}^{(3)}= \tilde\alpha_c T\sum_{\omega_{n}}\int
d\mathbf{q}\frac{\left|\omega_{n}\right|}{q}\left|
    \tilde{\varphi}\left(\mathbf{q},\omega_{n}\right)\right|^{2}.
\label{eq:S_dyn_z3}
\end{equation}

The action defined in Eqs.\ (\ref{eq:total_action}) to (\ref{eq:S_dyn_z3}) describes a
system close to a \emph{quantum} phase transition. In order to investigate a droplet in a
system close to a \emph{classical thermal} phase transition, we simply drop the dynamical
piece of the action and eliminate the imaginary time-dependence of the order parameter
field.

\section{Existence of magnetic droplets\label{sec:Droplet-static-profile}}

\begin{figure}
\includegraphics[width=6cm]{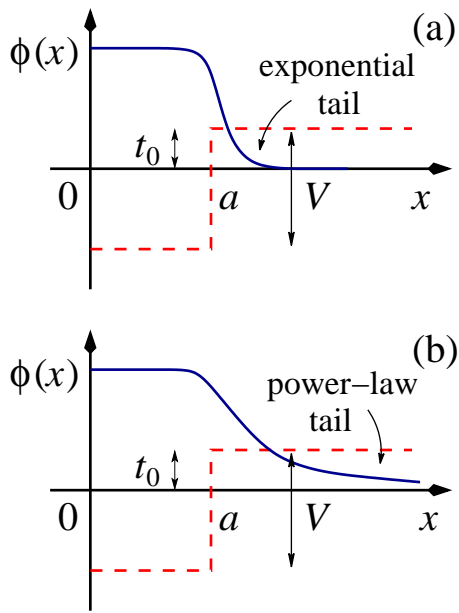}
\caption{(Color online) Schematic local order parameter profiles for defect induced
         droplets for short-range (a) and long-range (b) interactions. The dashed line
         depicts the defect potential.}
\label{fig:profile}
\end{figure}

In this section we investigate to what extent the existence of droplets is influenced by
the long-range spatial interaction. The basic idea is as follows: If the local potential
$t_0-V$ on the defect is negative, magnetic order is preferred on the defect even though
the bulk system may be nonmagnetic, $t_0>0$. Figure \ref{fig:profile} shows a schematic
of the local order parameter profile in this situation, comparing short-range and
long-range interactions.
In the short-range case, the tail of the droplet profile falls off exponentially outside
the defect.\cite{NVBK99b,MillisMorrSchmalian01,MillisMorrSchmalian02} Thus the tail
provides only a subleading surface term to the droplet free energy. In contrast, for the
long-range interaction (\ref{eq:I_kernel}), the tail must take a power-law form because
Griffiths theorem\cite{Griffiths67} dictates that the magnetic correlations cannot decay
faster than the interaction. The tail thus extends far into the bulk region where the
local potential is positive, and therefore leads to a large positive contribution to the
droplet free energy. In this section we study whether this mechanism hinders the
formation of magnetic droplets for long-range interactions.

\subsection{Saddle-point equation\label{subsec:SP-equation}}

We start by analyzing the action (\ref{eq:total_action}) within saddle-point
approximation, focusing on the case $t_0>0$ (noncritical bulk system) because it is
relevant for Griffiths phenomena. We can restrict ourselves to time-independent solutions
because they have the lowest saddle-point actions (any time dependence produces an extra,
strictly positive contribution from $S_{\rm dyn}$). Setting
$\varphi(\mathbf{x},\tau)=\phi(\mathbf{x})$ and minimizing the total action with respect
to this field leads to the saddle-point equation
\begin{equation}
\left(t_0 +\delta t\left(\mathbf{x}\right)+\Gamma_{0}\right)\phi\left(\mathbf{x}\right)+u\phi^{3}
\left(\mathbf{x}\right) = \! \int \! \frac{\gamma\phi\left(\mathbf{y}\right)d\mathbf{y}}
{\left[\xi_{0}^{2}+\left|\mathbf{x}-\mathbf{y}\right|^{2}\right]^{\frac{d+\sigma}{2}}}.
\label{eq:Saddle-point_stat}
\end{equation}
Note that the classical action discussed at the end of section \ref{sec:The-Model} leads
to the same saddle-point equation. Therefore, the remainder of this section applies to
both the classical and quantum cases.

We have not managed to solve the nonlinear integral equation (\ref{eq:Saddle-point_stat})
in closed form. We therefore first present analytical results for the behavior of
$\phi(\mathbf{x})$ far away from the defect, and then we complement them by a numerical
solution. For sufficiently large $V$ (such that $t_0-V$ is sufficiently negative), we
expect the order parameter in the droplet core, $|\mathbf{x}|<a$, to be roughly constant.
Griffiths' theorem\cite{Griffiths67} mandates that the droplet tail cannot decay faster
than $|\mathbf{x}|^{-(d+\sigma)}$; we therefore try the spherically symmetric ansatz
\begin{equation}
\phi\left(\mathbf{x}\right)= \left\{ \begin{array}{lr} \phi_{0} & \quad
(|\mathbf{x}|<a)\\ C/|\mathbf{x}|^{d+\sigma} & (|\mathbf{x}|> a)\end{array}\right. ~,
\label{eq:ansatz_phi}
\end{equation}
with parameters $\phi_0$ and $C$. Note that in general, the ansatz (\ref{eq:ansatz_phi})
is not continuous at $|\mathbf{x}|= a$. To cure this unphysical behavior, there must be an
intermediate region $a<|\mathbf{x}|<a+\xi_m$ which connects the core with the asymptotic region
in (\ref{eq:ansatz_phi}). We will come back to this point later in this
section.

We now insert the ansatz (\ref{eq:ansatz_phi}) into the saddle-point equation
(\ref{eq:Saddle-point_stat}) and analyze it in the limit of large defects, $a \gg \xi_0$,
and large distance, $|\mathbf{x}| \gg a$ where (\ref{eq:Saddle-point_stat}) can be
linearized in $\phi$. We find that the ansatz indeed solves the linearized saddle-point
equation with the amplitude $C$ given by (to leading order in $a$)
\begin{equation}
C=\frac{\Omega_d\phi_{0}\gamma}{dt_{0}} a^{d}~.
\label{eq:C-scaling}
\end{equation}
Note that $C$ diverges when the bulk system approaches criticality ($t_0 \to 0$)
indicating that the ansatz (\ref{eq:ansatz_phi}) is not valid for a defect in a
\emph{critical} bulk. We will come back to this point in the next subsection.

To determine $\phi_0$, we now calculate the saddle-point action by inserting the solution
(\ref{eq:ansatz_phi}) with (\ref{eq:C-scaling}) into the action (\ref{eq:total_action}).
The result is the sum of a droplet core term, a tail term, and a core-tail interaction
term. The core term takes the form $(\Omega_d/d)a^d\phi_0^2 (t_0-V +u\phi_0^2/2)$. The
contribution of the long-range interaction is exactly cancelled by the $\Gamma_0$-term,
as must be the case for a constant order parameter. Interestingly, the tail term and the
core-tail interaction term are subleading in the limit of large defects, $a \gg \xi_0$.
Their leading $a$-dependencies are $a^{d-2\sigma}$ and $a^{d-\sigma-1}$ (up to possible
logarithmic corrections), respectively. Finally, we have to consider the intermediate
region $a<|\mathbf{x}|<a+\xi_m$ in which droplet core smoothly connects to the asymptotic
tail. From the numerical solution of the saddle-point equation (discussed in the next
subsection) we found that the width of the intermediate region is of the order of the
microscopic scale, $\xi_m \sim \xi_0$ (at least as long as the bulk system is not too
close to criticality; see next subsection for details). Importantly, $\xi_m$ does not
depend on the defect size $a$. Therefore, the intermediate region can at most make a
surface-type contribution to the droplet free energy, i.e., it can at most scale like
$a^{d-1}$.

Collecting all the terms, we find that the saddle-point action takes the form
\begin{equation}
S_{\rm SP}=\frac{\Omega_d}{d}\phi_{0}^{2} a^{d}\left(t_{0}-V+\frac{u}{2}\phi_{0}^{2}\right)+{\cal O}\left(a^{d-1},a^{d-2\sigma}\right)
\label{eq:SP-action}
\end{equation}
in the limit of a large defect ($a\to \infty$). Minimizing $S_{\rm SP}$ with respect to $\phi_0$ gives the
optimal value
\begin{equation}
\phi_{0}=\sqrt{\frac{V-t_{0}}{u}}~.
\label{eq:phi_0-scaling}
\end{equation}
This means, in the limit of a large defect, a droplet of local order starts to form as
soon as the local potential $t_0-V$ on the defect becomes negative. For finite $a$, the
subleading terms in (\ref{eq:SP-action}) lead to a shift in the onset of local order that
can be described by finite-size scaling in the usual way.

The results (\ref{eq:SP-action}) and (\ref{eq:phi_0-scaling}) are identical to the case
of short-range interactions.\cite{NVBK99b,MillisMorrSchmalian01,MillisMorrSchmalian02} We
thus arrive at the somewhat surprising conclusion that even though the long-range
interactions do induce a power-law tail of the droplet, they do not change the leading
behavior of its free energy (in the limit of large defects), and thus do not hinder the
existence of large droplets.

We also note that an additional short-range interaction of the form
$\left|\nabla\phi\right|^{2}$ in the static action (\ref{eq:S_stat}) will not modify our
results. Clearly, in the core region of the droplet it plays no role, and faraway from
the core $\left(\mathbf{x} \gg a\right)$, it only produces a subleading power-law. Its
contribution in the intermediate region can at most be of order $a^{d-1}$.

\subsection{Fluctuations}
\label{subsec:Fluctuations}

So far, we have analyzed the magnetic droplets within saddle-point approximation. In this
subsection we discuss to what extent fluctuations modify the above saddle-point analysis.
It is useful to divide the fluctuations into two classes, small fluctuations about the
saddle-point solution and collective reorientations of the entire droplet in (imaginary)
time. These two classes are well separated if the local order on the defect is properly
developed, i.e., $V-t_0\gtrsim u$. The collective reorientations determine the long-time
quantum dynamics of the droplet. They will be considered in more detail in Sec.\
\ref{sec:The-dynamics}.

In contrast, small long-wavelength fluctuations could potentially modify the droplet
profile (\ref{eq:ansatz_phi}), in particular the form of the magnetization tail. To study
the relevance of these fluctuations, we expand the action (\ref{eq:total_action}) about
the saddle-point solution and perform a tree-level (power-counting) renormalization group
analysis. The results depend qualitatively on whether or not the bulk system is critical.

As long as the bulk system is in its disordered phase, $t_0>0$, the asymptotic
long-distance decay of the droplet magnetization is controlled by the \emph {stable}
large-$t_0$ fixed point of the bulk rather than its critical fixed point. Since this
stable fixed point does not have anomalous dimensions, the saddle-point analysis is
qualitatively correct and the decay exponent in (\ref{eq:ansatz_phi}) remains unchanged.
Thus, the fluctuations only renormalize nonuniversal prefactors. Analogous results were
found in Ref.\ \onlinecite{MillisMorrSchmalian01} for the case of short-range case
interaction. Note that critical fluctuations \emph{on the defect} can change the exponent
in the relation (\ref{eq:phi_0-scaling}) close to the onset of local order at $t_0-V=0$,
provided the system is below its upper critical dimension. However, this has no bearing
on the form of the tail.

In contrast, if the bulk system is right at the transition, $t_0=0$, the long-distance
magnetization decay is controlled by the exponent $\eta$ of the \emph{critical} fixed
point via $\phi(\mathbf{x}) \sim |\mathbf{x}|^{-d+2-\eta}$ (because far from the defect,
$\phi(\mathbf{x})$ falls off as the bulk correlation function). For a classical magnet
with long-range interactions this fixed point was studied in the seminal work of Fisher,
Ma, and Nickel.\cite{FisherMaNickel72} They found that the critical behavior is
mean-field-like for $\sigma<d/2$ with $\eta=2-\sigma$. For $\sigma>2-\eta_{\rm SR}$
(where $\eta_{\rm SR}$ is the exponent of the corresponding short-range model), the
critical behavior is identical to that of the short-range
model.\cite{Sak73,LuijtenBlote02,Cardy_book96} In between, the exponents are nonclassical
and interpolate between mean-field and short-range behavior. Lets us also point out that
interesting crossover phenomena occur when the bulk system is close but not exactly at
the critical point. In this case the critical fixed point controls the magnetization
decay at intermediate distances (of the order of the bulk correlation length) from the
defect while the asymptotic behavior is again given by the saddle-point result
(\ref{eq:ansatz_phi}).

\subsection{Numerical solutions of the saddle-point equation}
\label{subsec:Numerics}

In this subsection we confirm and complement the asymptotic analysis of the saddle-point
equation (\ref{eq:Saddle-point_stat}) by a numerically exact solution.

We study both one and three space dimensions. In the three-dimensional case, for a
spherical defect and droplet, the angular integration on the r.h.s. of the saddle-point
equation (\ref{eq:Saddle-point_stat}) can be carried out analytically leading to a
one-dimensional integral equation in radial direction. We now discretize space in units
of the microscopic length $\xi_{0}$ and fix the energy scale by setting $u=1$. The
resulting set of nonlinear equations is solved by the following procedure: We start from
an ansatz for $\phi$ (e.g., the ansatz given in (\ref{eq:ansatz_phi})) and numerically
perform the integral in the long-range term of (\ref{eq:Saddle-point_stat}). We then
determine an improved value for $\phi$ by solving the remaining cubic equation at each
point by standard methods. These steps are repeated iteratively until the solution
converges.

In this way, we have analyzed one-dimensional systems with $2\times 10^{4}$ to $2\times
10^{5}$ points and three-dimensional systems with $10^{4}$ to $10^{5}$ points in radial
direction. We have studied the cases $\sigma=1,2,3$, large defects $a \gg 1$ and various
values of  $t_0$, $V$ and $\gamma$. For weak long-range interactions and away from bulk
criticality, our procedure converges rapidly. With increasing $\gamma$ and decreasing
$t_0$, the convergence becomes slower. However, in all cases, our self-consistency cycle
eventually converges, giving us a numerically exact solution of the saddle-point
equation.

\begin{figure}
\begin{center}\includegraphics[width=7.2cm]{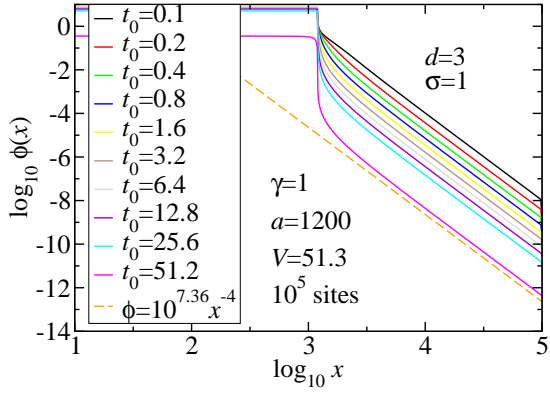}\end{center}
\caption{(Color online) Local order parameter $\phi$ of a three-dimensional droplet
          as a function of distance $x$ from the defect center for different distances
          $t_0=0.1$ to 51.2 from bulk criticality (from top to bottom).}
\label{fig:droplet-profile-3d}
\end{figure}

\begin{figure}
\begin{center}\includegraphics[width=6.8cm]{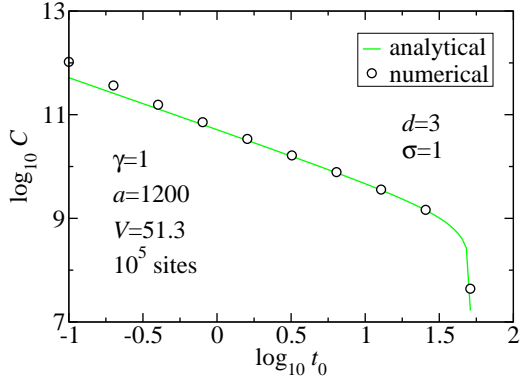}\end{center}
\caption{(Color online) Amplitude $C$ of the asymptotic power-law decay of the droplet tail
for the system shown in Fig.\ \ref{fig:droplet-profile-3d}. The solid line is the theoretical
prediction, Eqs.\ (\ref{eq:C-scaling}) and (\ref{eq:phi_0-scaling}). }
\label{fig:prefactor-3d}
\end{figure}

We now present and discuss a few characteristic results from these calculations.
In Fig.\ \ref{fig:droplet-profile-3d}, we show saddle-point solutions for $d=3$,
$\sigma=1$ and different values of the distance $t_0$ from bulk criticality.
In agreement with the analytical predictions of the last subsection, the order parameter
is essentially constant on the defect. For large $|\mathbf{x}|$, the droplet tail falls
off with the predicted power-law $\phi \sim |\mathbf{x}|^{-(d+\sigma)}= |\mathbf{x}|^{-4}$
for all values of $t_0$. The amplitude $C$ of this power-law decay is analyzed in Fig.\
\ref{fig:prefactor-3d}.
As predicted in the last subsection, for small $t_0$, $C$ behaves like $1/t_0$
(the small deviations are the lowest $t_0$ stem from the fact that in these cases,
$10^5$ sites is not sufficient to reach the asymptotic regime).

\begin{figure}
\begin{center}\includegraphics[width=7.2cm]{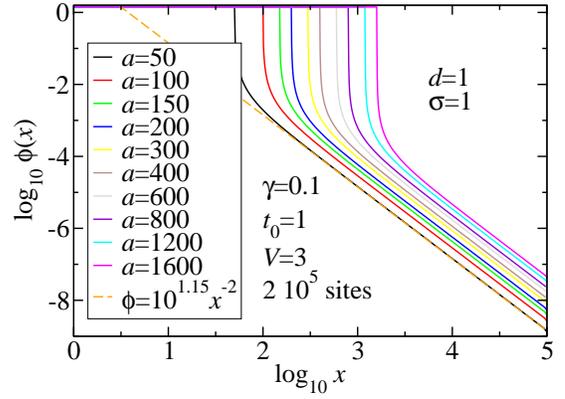}\end{center}
\caption{(Color online) Local order parameter $\phi$ of a one-dimensional droplet
          as a function of distance $x$ from the defect center for different defect
          sizes $a=50$ to 1600 (from left to right).}
\label{fig:droplet-profile-1d}
\end{figure}

\begin{figure}
\begin{center}\includegraphics[width=6.8cm]{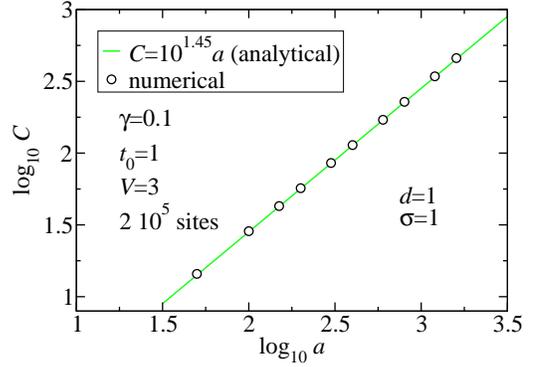}\end{center}
\caption{(Color online) Amplitude $C$ of the asymptotic power-law decay of the droplet tail
for the system shown in Fig.\ \ref{fig:droplet-profile-1d}. The solid line is the theoretical
prediction, Eqs.\ (\ref{eq:C-scaling}) and (\ref{eq:phi_0-scaling}). }
\label{fig:prefactor-1d}
\end{figure}

Figure \ref{fig:droplet-profile-1d} shows the dependence of the droplet profile on the
size $a$ of the defect for a system with $d=\sigma=1$.
For all $a$, the asymptotic decay of the droplet tail takes the predicted power-law form,
$\phi \sim |\mathbf{x}|^{-(d+\sigma)}= |\mathbf{x}|^{-2}$. This figure also shows that
the width $\xi_m$ of the intermediate $\mathbf{x}$-region which connects the droplet core
with the power-law tail does not change with $a$ as discussed in the last subsection.
(This becomes even more obvious when a linear rather than the logarithmic $x$-scale is
used.) Moreover, in agreement with (\ref{eq:phi_0-scaling}), $\phi_{0}$ does not depend
on $a$. The amplitude $C$ of this power-law decay is analyzed in Fig.\
\ref{fig:prefactor-1d}.
In agreement with the theoretical prediction (\ref{eq:C-scaling}), the amplitude grows
linearly with the defect size $a$.

We have performed analogous calculations for other parameter sets, including varying
$t_0$ for $a=300,~600, ~1600$ as well as varying $a$ for $V=30$. In all cases, we have found
excellent agreement with the predictions of the asymptotic analysis of Sec.\
\ref{subsec:Numerics}.

\section{Tunneling dynamics\label{sec:The-dynamics}}

In this section, we study the tunneling dynamics of a single droplet at a
zero-temperature quantum phase transition. Our approach starts from the pioneering work
of Callan and Coleman\cite{CallanColeman77} and Leggett and
coworkers\cite{CaldeiraLeggett83,LCDFGZ87} (in the case of dissipative dynamics). In the
following subsections we separately discuss the droplet dynamics for the three dynamical
actions given in Eqs.\ (\ref{eq:S_dyn_z1}) to (\ref{eq:S_dyn_z3}), starting with the
undamped case.

\subsection{Undamped magnet}
\label{subsec:undamped}

Following Callan and Coleman\cite{CallanColeman77}, the tunneling rate between the
\emph{up} and \emph{down} states of the droplet (i.e., the tunnel splitting of the ground
state) can be estimated from the action of instanton-like saddle-point solutions
$\varphi(\mathbf{x},\tau)$ fulfilling the boundary conditions $\varphi(\mathbf{x},\tau)
\to \pm \phi(\mathbf{x})$ for $\tau \to \pm \infty$. In principle, several processes
contribute to the overall tunneling rate. In the simplest one, the droplet retains its
shape while collapsing and reforming.\cite{BeanLivingston59,ChudnovskyGunther88} A
competing process consists of the nucleation of a domain wall that then sweeps the
droplet.\cite{Stauffer76}

We start by considering the collapse-and-reformation process which can be described by an
ansatz
\begin{equation}
\varphi\left(\mathbf{x},\tau\right)=\phi\left(\mathbf{x}\right)\eta\left(\tau\right)
\label{eq:phi(x)eta(t)}
\end{equation}
with $\phi(\mathbf{x})$ being the static saddle-point solution of section
\ref{sec:Droplet-static-profile} and $\eta(\tau) \to \pm 1$ for $\tau \to \pm \infty$.
Inserting this ansatz into the action (\ref{eq:total_action}) and integrating over the
spatial variables yields the following excess effective action (above the
time-independent solution $\eta\equiv 1$)
\begin{equation}
\Delta S^{(1)}=\frac{\Omega_d}{d}\phi_{0}^{2}a^{d}\int d\tau\left[ \frac{\phi_0^2
u}{2}\left(1-\eta^{2}\right)^2+
   \frac{1}{c^{2}}\left(\frac{d\eta}{d\tau}\right)^{2}\right]
\label{eq:S_eff_z1}
\end{equation}
to leading order in the defect size $a$. The saddle-point instanton solution of this action
can be found exactly. It takes the form $\eta\left(\tau\right)=\tanh\left(\tau/\tau_0\right)$, with
$\tau_0^{-2}=c^2\phi_0^2u/2$. The resulting instanton action reads
\begin{equation}
\Delta S_{\rm inst}^{(1)}=\frac {4\Omega_d}{3d} u\phi_{0}^{4}a^{d}\tau_0
\label{eq:Sinst_1}
\end{equation}
giving a tunnel splitting
\begin{equation}
\omega^{(1)}\approx \omega_0 e^{-\Delta S_{\rm inst}^{(1)}}.
 \label{eq:rate_1}
\end{equation}
The ``attempt frequency'' $\omega_0$ can be determined by standard quantum tunneling
considerations\cite{LCDFGZ87} from the fluctuations about the instanton solution.
Importantly, the tunneling rate decays exponentially with the volume of the droplet.
Equations (\ref{eq:S_eff_z1}) to (\ref{eq:rate_1}) are in complete agreement with the
corresponding results for the case of short-range
interactions,\cite{MillisMorrSchmalian01,MillisMorrSchmalian02,HoyosVojta_unpublished}
reflecting the fact that the leading terms of the instanton action stem from the droplet
core rather than the tail.

To discuss the contribution of the moving domain wall processes to the tunneling rate,
we use the ansatz
\begin{equation}
\varphi(\mathbf{x},\tau)=\phi(\mathbf{x})\eta(\tau-x/v)
\label{eq:phi(x)eta(t-xv)}
\end{equation}
which describes a domain wall that sweeps the droplet in $x$-direction with
velocity $v$. $\eta$ describes the domain wall shape and fulfills the boundary conditions
$\eta(z) \to \pm 1$ for $z \to \pm \infty$ as before. Inserting this into the
action (\ref{eq:total_action}) gives the same effective action (\ref{eq:S_eff_z1}) plus
one additional positive term from the spatial dependence of $\eta$ (this term corresponds
to the domain wall energy). Therefore, the minimal action for this process is bounded by
(\ref{eq:Sinst_1}), and to exponential accuracy the corresponding tunnelling rate cannot
be larger than (\ref{eq:rate_1}). This is in qualitative agreement with earlier results
for short-range interactions. Stauffer\cite{Stauffer76} estimated the tunneling rate of a
domain wall within quasiclassical WKB approximation and found that it depends exponentially
on the droplet volume. Senthil and Sachdev\cite{SenthilSachdev96} estimated the tunnel
splitting of a locally ordered island in a transverse-field Ising model using perturbation
arguments. Again, the result (which should contain all possible processes) is
exponentially small in the droplet volume.

\subsection{Overdamped dynamics}
\label{subsec:overdamped}

We now consider overdamped dynamics with the action $S_{\rm dyn}^{(2)}$ as given in
(\ref{eq:S_dyn_z2}).  Inserting the ansatz $\varphi\left(\mathbf{x},\tau\right)=
\phi\left(\mathbf{x}\right)\eta\left(\tau\right)$
into $S_{\rm dyn}^{(2)}$ and integrating over the spatial variables gives the following
contribution to the effective action
\begin{equation}
\Delta S^{(2)}= \frac {\Omega_d}{d} a^d \phi_0^2
 \int d\tau d\tau^{\prime} \frac{\tilde\alpha}{2\pi}
            \frac{(\eta(\tau)-\eta(\tau^{\prime}))^2}{\left(\tau-\tau^{\prime}\right)^{2}}
\label{eq:S_eff_z2}
\end{equation}
to leading order in the defect size $a$. The other terms are as in eq.\
(\ref{eq:S_eff_z1}). A straight forward saddle-point instanton analysis of the
effective action analogous to the last subsection fails because the interaction
of the trajectory $\eta(\tau)$ at large positive times with the trajectory at
large negative times causes a logarithmic divergence. Following Refs.\
\onlinecite{LCDFGZ87} and \onlinecite{DorseyFisherWartak86}, the calculation
therefore proceeds in two stages.

In the first stage, we introduce a \emph{low-frequency} cutoff $\omega_c$ in the dynamic
action (\ref{eq:S_dyn_z2}). This changes the interaction kernel in (\ref{eq:S_eff_z2}),
\begin{equation}
\frac{1}{\left(\tau-\tau^{\prime}\right)^{2}} \to
\frac{1+2\omega_c|\tau-\tau^{\prime}|}{\left(\tau-\tau^{\prime}\right)^{2}(1+\omega_c|\tau-\tau^{\prime}|)^2}~,
\end{equation}
and removes the divergence. We have not been able to solve analytically for the instanton
but we have used the ansatz $\eta(\tau)=\tanh(\tau/\tau_0)$ with variational parameter
$\tau_0$. Minimizing the effective action $\Delta S^{(1)}+ \Delta S^{(2)}$ with respect
to $\tau_0$ gives
\begin{equation}
\tau_0 = \frac {3 \tilde \alpha}{\pi \phi_0^2 u} \left(1+ \sqrt{1+\frac{2 \pi^2 \phi_0^2
u}{9 c^2 \tilde\alpha^2}} \right)~.
\end{equation}
In the limit of weak dissipation, $\tilde\alpha \to 0$, we recover the result for
undamped dynamics while strong dissipation, $\tilde\alpha\to \infty$, leads to $\tau_0 =
6\tilde\alpha /(\pi\phi_0^2 u)$. The resulting instanton action can be expressed in terms
of the dimensionless dissipation strength parameter \cite{LCDFGZ87,DorseyFisherWartak86}
\begin{equation}
\alpha=\frac {4\Omega_d}{\pi d} \phi_{0}^{2}a^{d} \tilde\alpha~.
\label{eq:alpha}
\end{equation}
We note that $\alpha$ is proportional to the defect volume $a^d$. (Analogous results have
been obtained for dissipative random quantum Ising
models.\cite{SchehrRieger06,HoyosVojta06}) The dissipative part of the instanton action
reads
\begin{equation}
\Delta S_{\rm inst}^{(2)}= -\alpha \ln(\omega_c) +f(\alpha) ~, \label{eq:Sinst_2}
\end{equation}
where the function $f(\alpha)$ is given by  $f(\alpha) = c\, \alpha +O(\alpha^2)$ for
weak dissipation and  $f(\alpha) = -\alpha \ln \alpha + c^{\prime}\, \alpha
+O(\ln(\alpha))$ for strong dissipation. $c$ and $c^{\prime}$ are constants of order one.
For comparison we have also studied a piecewise linear ansatz for $\eta(\tau)$. The
resulting instanton action is identical to (\ref{eq:Sinst_2}) except for different
numerical values of the constants $c,c^\prime$. At the end of the first stage of the
calculation we thus obtain the bare tunnel splitting
\begin{equation}
\omega_{\rm bare}^{(2)}\approx \omega_0 e^{-\Delta S_{\rm inst}^{(2)}}~.
\label{eq:bare_rate_2}
\end{equation}

In the second stage of the calculation the resulting dissipative two-level system
is treated using renormalization group methods. \cite{LCDFGZ87} It is well-known
that instanton-instanton interactions renormalize the tunnel splitting, yielding
\begin{equation}
\omega^{(2)} \sim \omega_{\rm bare}^{(2)}\left[\frac{\omega_{\rm bare}^{(2)}}{\omega_c}
\right]^{\alpha/(1-\alpha)}.
\label{eq:rate_2}
\end{equation}
This also eliminates the unphysical dependence of the tunnel splitting on the arbitrary
cutoff parameter $\omega_c$. We thus find that the smaller defects with $\alpha<1$
continue to tunnel, albeit with a strongly reduced rate. The larger defects with
$\alpha>1$ cease to tunnel, i.e., they are on the localized side of the
Kosterlitz-Thouless phase transition of the dissipative two-level system. These results
are in qualitative agreement with the case of short-range
interactions.\cite{MillisMorrSchmalian01,MillisMorrSchmalian02}

\subsection{Conserved overdamped dynamics}
\label{subsec:ferro}

Finally, we consider the case of overdamped dynamics with a conserved order parameter
as given by the dynamic action (\ref{eq:S_dyn_z3}). Such an action arises,
e.g., in the case of an itinerant quantum ferromagnet.

Order parameter conservation requires some care in discussing the dynamics of our locally
ordered droplet. In particular, the homogeneous magnetization $\int d\mathbf{x}
\varphi(\mathbf{x},\tau)$ must not be time dependent. Therefore, the product form
$\varphi\left(\mathbf{x},\tau\right)= \phi\left(\mathbf{x}\right)\eta\left(\tau\right)$
with $\phi(\mathbf{x})$ the static solution of section \ref{sec:Droplet-static-profile}
is not a suitable ansatz in this case. This can be fixed (in a crude way) by subtracting
a constant from the droplet profile, $\phi^\prime(\mathbf{x})= \phi(\mathbf{x}) -{\rm
const}$ such that the $\mathbf{q}=0$ Fourier component is cancelled. The ansatz
$\varphi\left(\mathbf{x},\tau\right)=
\phi^\prime\left(\mathbf{x}\right)\eta\left(\tau\right)$ then provides a variational
upper bound for the instanton action.

Inserting this ansatz into (\ref{eq:S_dyn_z3}) and carrying out the integral over the
spatial variables leads to a dissipative term in the effective $\eta(\tau)$ action with
the same functional form as (\ref{eq:S_eff_z2}). The prefactor and the resulting
dimensionless dissipation strength $\alpha$, however, are different. To leading order in
the defect size $a$, we find
\begin{equation}
\alpha = \left\{\begin{array}{cc} 8\phi_0^2 a^4 \tilde\alpha_c/\pi  & \quad (d=3)
\\ 32 \phi_0^2 a^3 \tilde\alpha_c /(3\pi)& \quad (d=2)
\end{array} \right.~.
\end{equation}
In general dimension $d\ge 2$, the dimensionless dissipation strength is now proportional
to $a^{d+1}$ instead $a^d$. The extra factor $a$ compared to the non-conserved case in
Sec.\ \ref{subsec:overdamped} can be understood as follows. To invert the magnetization
of a droplet of linear size $a$, magnetization must be transported over a distance that
is at least of order $a$ (because the order parameter conservation prevents simply
flipping the sign of the magnetization on the defect). This involves modes with wave
vectors of the order of $q\sim 1/a$. Since the dissipation strength in
(\ref{eq:S_dyn_z3}) is inversely proportional to $q$, we expect an additional factor $a$
in the effective action. This argument strongly suggests that this extra factor is
\emph{not} an artefact of our simple ansatz for $\varphi\left(\mathbf{x},\tau\right)$ but
correctly reflects the physics of the conserved order parameter case.

In all other respects, the calculation proceeds as in the non-conserved case in
Sec.\ \ref{subsec:overdamped}. The resulting dynamic behavior of the droplets depends
on the value of the dimensionless dissipation strength parameter $\alpha$.
Small droplets ($\alpha<1$) still tunnel while the larger ones ($\alpha>1$) freeze.
Because $\alpha$ is now proportional to $a^{d+1}$ the tunneling of large droplets is
even more strongly suppressed than in the non-conserved case.

\section{Discussion and conclusions\label{sec:conclusions}}

To summarize, we have studied the physics of a single defect coupling to the square of
the order parameter in a nearly critical system with long-range spatial interactions of
the form $r^{-(d+\sigma)}$ with $\sigma>0$. Such a defect can induce the nucleation of a
magnetic droplet while the bulk system is still in the nonmagnetic phase. Due to the
long-range interactions, the droplet magnetization develops a long power-law tail, i.e.,
at large distances $r$ from the defect, it decays like $r^{-(d+\sigma)}$ in agreement
with Griffiths' theorem.\cite{Griffiths67} Nonetheless, the droplet free energy is
dominated by the core (on-defect) contribution while the tail contribution is subleading
in the limit of large defects. Therefore, droplets will nucleate on large defects as soon
as the local potential (the local distance from criticality) becomes negative, in
complete agreement with the case of short-range interactions. Our explicit calculations
of the droplet magnetization profile have been performed within saddle-point
approximation, but as long as the bulk system is noncritical, fluctuations do not change
the functional form of the droplet. They only renormalize nonuniversal parameters.

In addition to the existence of the magnetic droplets, we have also investigated their
dynamics. As is well known,\cite{Vojta06} in the case of a classical (thermal) phase
transition, the droplet cannot order statically. Instead, it fluctuates between `up' and
`down' due to thermal fluctuations. For a zero-temperature quantum phase transition, the
behavior is potentially different, depending on the form of the dynamic action. We have
studied three cases. In the absence of dissipation, even very large droplets can always
tunnel, but with a rate that decreases exponentially with the droplet volume. This
changes in the presence of (Ohmic) dissipation. The qualitative behavior now depends on
the dimensionless dissipation strength $\alpha$. For $\alpha<1$, the droplet still
tunnels albeit with a further reduced rate while for $\alpha>1$, tunneling ceases and the
droplet magnetization becomes static. For overdamped dynamics without order parameter
conservation, $\alpha$ is proportional to the volume of the droplet core. Thus,
sufficiently large droplets always freeze in agreement with Refs.\
\onlinecite{CastroNetoJones00,MillisMorrSchmalian01,MillisMorrSchmalian02}. In the case
of overdamped dynamics with order parameter conservation as in the itinerant quantum
ferromagnet, the dissipation effects are further enhanced because the
dimensionless dissipation strength $\alpha$ for a droplet of linear core size $a$ is
proportional to $a^{d+1}$ rather than $a^d$.

Let us comment on the order parameter symmetry. Our explicit results have been for the
case of a scalar (Ising) order parameter. However, the analysis of the droplet existence
in Sec.\ \ref{sec:Droplet-static-profile} relied on saddle-point arguments and thus
applies equally to continuous $O(N)$ order parameters with $N>1$. In contrast, to
generalize the discussion of the dynamics in Sec.\ \ref{sec:The-dynamics} to such order
parameters, other types of fluctuations (rotational ones) must be considered.

We also emphasize that we have discussed the case of an isotropic attractive long-range
interaction. Droplet \emph{formation} dominated by oscillating and/or anisotropic
interactions such as the dipolar or the RKKY interactions is likely of different type and
not considered here.

Finally, we briefly discuss the consequences of our results for the (quantum) Griffiths
effects in systems with long-range spatial interactions. Because the power-law
magnetization tail only makes a subleading contribution to the free energy of a magnetic
droplet, such droplets can form on rare (strongly coupled) spatial regions of the
disordered system essentially in the same way as in the case of short-range interactions.
Therefore, as long as droplet-droplet coupling can be neglected, the Griffiths effects
should be identical to those in short-range interacting systems. However, it is clear
that the droplet-droplet coupling is more important for long-range interactions than for
short-range ones. This means, it must be considered for lower droplet density and, in the
quantum case, for higher temperatures. The complicated physics caused by the coupling of
several droplets is beyond the scope of this paper. Recently, it has been argued
\cite{DobrosavljevicMiranda05} that this coupling can qualitatively change the Griffiths
effects at least in some cases. A complete understanding of this phenomenon remains a
task for the future.

\begin{acknowledgments}
We gratefully acknowledge discussions with J.\ Schmalian and M.\ Vojta. This work has
been supported by the NSF under grant no. DMR-0339147, and by Research Corporation. We
also thank the Aspen Center for Physics, where part of this work has been performed.

\end{acknowledgments}

\bibliographystyle{apsrev}
\bibliography{../00Bibtex/rareregions}

\begin{thebibliography}{41}
\expandafter\ifx\csname natexlab\endcsname\relax\def\natexlab#1{#1}\fi
\expandafter\ifx\csname bibnamefont\endcsname\relax
  \def\bibnamefont#1{#1}\fi
\expandafter\ifx\csname bibfnamefont\endcsname\relax
  \def\bibfnamefont#1{#1}\fi
\expandafter\ifx\csname citenamefont\endcsname\relax
  \def\citenamefont#1{#1}\fi
\expandafter\ifx\csname url\endcsname\relax
  \def\url#1{\texttt{#1}}\fi
\expandafter\ifx\csname urlprefix\endcsname\relax\def\urlprefix{URL }\fi
\providecommand{\bibinfo}[2]{#2}
\providecommand{\eprint}[2][]{\url{#2}}

\bibitem[{\citenamefont{Griffiths}(1969)}]{Griffiths69}
\bibinfo{author}{\bibfnamefont{R.~B.} \bibnamefont{Griffiths}},
  \bibinfo{journal}{Phys. Rev. Lett.} \textbf{\bibinfo{volume}{23}},
  \bibinfo{pages}{17} (\bibinfo{year}{1969}).

\bibitem[{\citenamefont{Randeria et~al.}(1985)\citenamefont{Randeria, Sethna,
  and Palmer}}]{RanderiaSethnaPalmer85}
\bibinfo{author}{\bibfnamefont{M.}~\bibnamefont{Randeria}},
  \bibinfo{author}{\bibfnamefont{J.~P.} \bibnamefont{Sethna}},
  \bibnamefont{and} \bibinfo{author}{\bibfnamefont{R.~G.}
  \bibnamefont{Palmer}}, \bibinfo{journal}{Phys. Rev. Lett.}
  \textbf{\bibinfo{volume}{54}}, \bibinfo{pages}{1321} (\bibinfo{year}{1985}).

\bibitem[{\citenamefont{Wortis}(1974)}]{Wortis74}
\bibinfo{author}{\bibfnamefont{M.}~\bibnamefont{Wortis}},
  \bibinfo{journal}{Phys. Rev. B} \textbf{\bibinfo{volume}{10}},
  \bibinfo{pages}{4665} (\bibinfo{year}{1974}).

\bibitem[{\citenamefont{Harris}(1975)}]{Harris75}
\bibinfo{author}{\bibfnamefont{A.~B.} \bibnamefont{Harris}},
  \bibinfo{journal}{Phys. Rev. B} \textbf{\bibinfo{volume}{12}},
  \bibinfo{pages}{203} (\bibinfo{year}{1975}).

\bibitem[{\citenamefont{Bray and Huifang}(1989)}]{BrayHuifang89}
\bibinfo{author}{\bibfnamefont{A.~J.} \bibnamefont{Bray}} \bibnamefont{and}
  \bibinfo{author}{\bibfnamefont{D.}~\bibnamefont{Huifang}},
  \bibinfo{journal}{Phys. Rev. B} \textbf{\bibinfo{volume}{40}},
  \bibinfo{pages}{6980} (\bibinfo{year}{1989}).

\bibitem[{\citenamefont{Imry}(1977)}]{Imry77}
\bibinfo{author}{\bibfnamefont{Y.}~\bibnamefont{Imry}}, \bibinfo{journal}{Phys.
  Rev. B} \textbf{\bibinfo{volume}{15}}, \bibinfo{pages}{4448}
  (\bibinfo{year}{1977}).

\bibitem[{\citenamefont{McCoy and Wu}(1968{\natexlab{a}})}]{McCoyWu68}
\bibinfo{author}{\bibfnamefont{B.~M.} \bibnamefont{McCoy}} \bibnamefont{and}
  \bibinfo{author}{\bibfnamefont{T.~T.} \bibnamefont{Wu}},
  \bibinfo{journal}{Phys. Rev. Lett.} \textbf{\bibinfo{volume}{21}},
  \bibinfo{pages}{549} (\bibinfo{year}{1968}{\natexlab{a}}).

\bibitem[{\citenamefont{McCoy and Wu}(1968{\natexlab{b}})}]{McCoyWu68a}
\bibinfo{author}{\bibfnamefont{B.~M.} \bibnamefont{McCoy}} \bibnamefont{and}
  \bibinfo{author}{\bibfnamefont{T.~T.} \bibnamefont{Wu}},
  \bibinfo{journal}{Phys. Rev.} \textbf{\bibinfo{volume}{176}},
  \bibinfo{pages}{631} (\bibinfo{year}{1968}{\natexlab{b}}).

\bibitem[{\citenamefont{Fisher}(1992)}]{Fisher92}
\bibinfo{author}{\bibfnamefont{D.~S.} \bibnamefont{Fisher}},
  \bibinfo{journal}{Phys. Rev. Lett.} \textbf{\bibinfo{volume}{69}},
  \bibinfo{pages}{534} (\bibinfo{year}{1992}).

\bibitem[{\citenamefont{Fisher}(1995)}]{Fisher95}
\bibinfo{author}{\bibfnamefont{D.~S.} \bibnamefont{Fisher}},
  \bibinfo{journal}{Phys. Rev. B} \textbf{\bibinfo{volume}{51}},
  \bibinfo{pages}{6411} (\bibinfo{year}{1995}).

\bibitem[{\citenamefont{Vojta}(2003{\natexlab{a}})}]{Vojta03b}
\bibinfo{author}{\bibfnamefont{T.}~\bibnamefont{Vojta}}, \bibinfo{journal}{J.
  Phys. A} \textbf{\bibinfo{volume}{36}}, \bibinfo{pages}{10921}
  (\bibinfo{year}{2003}{\natexlab{a}}).

\bibitem[{\citenamefont{Sknepnek and Vojta}(2004)}]{SknepnekVojta04}
\bibinfo{author}{\bibfnamefont{R.}~\bibnamefont{Sknepnek}} \bibnamefont{and}
  \bibinfo{author}{\bibfnamefont{T.}~\bibnamefont{Vojta}},
  \bibinfo{journal}{Phys. Rev. B} \textbf{\bibinfo{volume}{69}},
  \bibinfo{pages}{174410} (\bibinfo{year}{2004}).

\bibitem[{\citenamefont{Berche et~al.}(1998)\citenamefont{Berche, Berche,
  Igloi, and Palagyi}}]{BBIP98}
\bibinfo{author}{\bibfnamefont{B.}~\bibnamefont{Berche}},
  \bibinfo{author}{\bibfnamefont{P.~E.} \bibnamefont{Berche}},
  \bibinfo{author}{\bibfnamefont{F.}~\bibnamefont{Igloi}}, \bibnamefont{and}
  \bibinfo{author}{\bibfnamefont{G.}~\bibnamefont{Palagyi}},
  \bibinfo{journal}{J. Phys. A} \textbf{\bibinfo{volume}{31}},
  \bibinfo{pages}{5193} (\bibinfo{year}{1998}).

\bibitem[{\citenamefont{Vojta}(2004)}]{Vojta04}
\bibinfo{author}{\bibfnamefont{T.}~\bibnamefont{Vojta}},
  \bibinfo{journal}{Phys. Rev. E} \textbf{\bibinfo{volume}{70}},
  \bibinfo{pages}{026108} (\bibinfo{year}{2004}).

\bibitem[{\citenamefont{Vojta}(2006)}]{Vojta06}
\bibinfo{author}{\bibfnamefont{T.}~\bibnamefont{Vojta}}, \bibinfo{journal}{J.
  Phys. A} \textbf{\bibinfo{volume}{39}}, \bibinfo{pages}{R143}
  (\bibinfo{year}{2006}).

\bibitem[{\citenamefont{Castro~Neto and Jones}(2000)}]{CastroNetoJones00}
\bibinfo{author}{\bibfnamefont{A.~H.} \bibnamefont{Castro~Neto}}
  \bibnamefont{and} \bibinfo{author}{\bibfnamefont{B.~A.} \bibnamefont{Jones}},
  \bibinfo{journal}{Phys. Rev. B} \textbf{\bibinfo{volume}{62}},
  \bibinfo{pages}{14975} (\bibinfo{year}{2000}).

\bibitem[{\citenamefont{Millis et~al.}(2001)\citenamefont{Millis, Morr, and
  Schmalian}}]{MillisMorrSchmalian01}
\bibinfo{author}{\bibfnamefont{A.~J.} \bibnamefont{Millis}},
  \bibinfo{author}{\bibfnamefont{D.~K.} \bibnamefont{Morr}}, \bibnamefont{and}
  \bibinfo{author}{\bibfnamefont{J.}~\bibnamefont{Schmalian}},
  \bibinfo{journal}{Phys. Rev. Lett.} \textbf{\bibinfo{volume}{87}},
  \bibinfo{pages}{167202} (\bibinfo{year}{2001}).

\bibitem[{\citenamefont{Millis et~al.}(2002)\citenamefont{Millis, Morr, and
  Schmalian}}]{MillisMorrSchmalian02}
\bibinfo{author}{\bibfnamefont{A.~J.} \bibnamefont{Millis}},
  \bibinfo{author}{\bibfnamefont{D.~K.} \bibnamefont{Morr}}, \bibnamefont{and}
  \bibinfo{author}{\bibfnamefont{J.}~\bibnamefont{Schmalian}},
  \bibinfo{journal}{Phys. Rev. B} \textbf{\bibinfo{volume}{66}},
  \bibinfo{pages}{174433} (\bibinfo{year}{2002}).

\bibitem[{\citenamefont{Vojta}(2003{\natexlab{b}})}]{Vojta03a}
\bibinfo{author}{\bibfnamefont{T.}~\bibnamefont{Vojta}},
  \bibinfo{journal}{Phys. Rev. Lett.} \textbf{\bibinfo{volume}{90}},
  \bibinfo{pages}{107202} (\bibinfo{year}{2003}{\natexlab{b}}).

\bibitem[{\citenamefont{Vojta et~al.}(1996)\citenamefont{Vojta, Belitz,
  Narayanan, and Kirkpatrick}}]{VBNK96}
\bibinfo{author}{\bibfnamefont{T.}~\bibnamefont{Vojta}},
  \bibinfo{author}{\bibfnamefont{D.}~\bibnamefont{Belitz}},
  \bibinfo{author}{\bibfnamefont{R.}~\bibnamefont{Narayanan}},
  \bibnamefont{and} \bibinfo{author}{\bibfnamefont{T.~R.}
  \bibnamefont{Kirkpatrick}}, \bibinfo{journal}{Europhys. Lett.}
  \textbf{\bibinfo{volume}{36}}, \bibinfo{pages}{191} (\bibinfo{year}{1996}).

\bibitem[{\citenamefont{Vojta et~al.}(1997)\citenamefont{Vojta, Belitz,
  Narayanan, and Kirkpatrick}}]{VBNK97}
\bibinfo{author}{\bibfnamefont{T.}~\bibnamefont{Vojta}},
  \bibinfo{author}{\bibfnamefont{D.}~\bibnamefont{Belitz}},
  \bibinfo{author}{\bibfnamefont{R.}~\bibnamefont{Narayanan}},
  \bibnamefont{and} \bibinfo{author}{\bibfnamefont{T.~R.}
  \bibnamefont{Kirkpatrick}}, \bibinfo{journal}{Z. Phys. B}
  \textbf{\bibinfo{volume}{103}}, \bibinfo{pages}{451} (\bibinfo{year}{1997}).

\bibitem[{\citenamefont{Belitz et~al.}(1997)\citenamefont{Belitz, Kirkpatrick,
  and Vojta}}]{BelitzKirkpatrickVojta97}
\bibinfo{author}{\bibfnamefont{D.}~\bibnamefont{Belitz}},
  \bibinfo{author}{\bibfnamefont{T.~R.} \bibnamefont{Kirkpatrick}},
  \bibnamefont{and} \bibinfo{author}{\bibfnamefont{T.}~\bibnamefont{Vojta}},
  \bibinfo{journal}{Phys. Rev. B} \textbf{\bibinfo{volume}{55}},
  \bibinfo{pages}{9452} (\bibinfo{year}{1997}).

\bibitem[{\citenamefont{Belitz et~al.}(2005)\citenamefont{Belitz, Kirkpatrick,
  and Vojta}}]{BelitzKirkpatrickVojta05}
\bibinfo{author}{\bibfnamefont{D.}~\bibnamefont{Belitz}},
  \bibinfo{author}{\bibfnamefont{T.~R.} \bibnamefont{Kirkpatrick}},
  \bibnamefont{and} \bibinfo{author}{\bibfnamefont{T.}~\bibnamefont{Vojta}},
  \bibinfo{journal}{Rev. Mod. Phys.} \textbf{\bibinfo{volume}{77}},
  \bibinfo{pages}{579} (\bibinfo{year}{2005}).

\bibitem[{\citenamefont{Griffiths}(1967)}]{Griffiths67}
\bibinfo{author}{\bibfnamefont{R.~B.} \bibnamefont{Griffiths}},
  \bibinfo{journal}{J. Math. Phys.} \textbf{\bibinfo{volume}{8}},
  \bibinfo{pages}{478} (\bibinfo{year}{1967}).

\bibitem[{\citenamefont{Narayanan et~al.}(1999)\citenamefont{Narayanan, Vojta,
  Belitz, and Kirkpatrick}}]{NVBK99b}
\bibinfo{author}{\bibfnamefont{R.}~\bibnamefont{Narayanan}},
  \bibinfo{author}{\bibfnamefont{T.}~\bibnamefont{Vojta}},
  \bibinfo{author}{\bibfnamefont{D.}~\bibnamefont{Belitz}}, \bibnamefont{and}
  \bibinfo{author}{\bibfnamefont{T.~R.} \bibnamefont{Kirkpatrick}},
  \bibinfo{journal}{Phys. Rev. B} \textbf{\bibinfo{volume}{60}},
  \bibinfo{pages}{10150} (\bibinfo{year}{1999}).

\bibitem[{\citenamefont{Fisher et~al.}(1972)\citenamefont{Fisher, Ma, and
  Nickel}}]{FisherMaNickel72}
\bibinfo{author}{\bibfnamefont{M.~E.} \bibnamefont{Fisher}},
  \bibinfo{author}{\bibfnamefont{S.-K.} \bibnamefont{Ma}}, \bibnamefont{and}
  \bibinfo{author}{\bibfnamefont{B.~G.} \bibnamefont{Nickel}},
  \bibinfo{journal}{Phys. Rev. Lett.} \textbf{\bibinfo{volume}{29}},
  \bibinfo{pages}{917} (\bibinfo{year}{1972}).

\bibitem[{\citenamefont{Sak}(1973)}]{Sak73}
\bibinfo{author}{\bibfnamefont{J.}~\bibnamefont{Sak}}, \bibinfo{journal}{Phys.
  Rev. B} \textbf{\bibinfo{volume}{8}}, \bibinfo{pages}{281}
  (\bibinfo{year}{1973}).

\bibitem[{\citenamefont{Luijten and Bl{\"o}te}(2002)}]{LuijtenBlote02}
\bibinfo{author}{\bibfnamefont{E.}~\bibnamefont{Luijten}} \bibnamefont{and}
  \bibinfo{author}{\bibfnamefont{H.~W.~J.} \bibnamefont{Bl{\"o}te}},
  \bibinfo{journal}{Phys. Rev. Lett} \textbf{\bibinfo{volume}{89}},
  \bibinfo{pages}{025703} (\bibinfo{year}{2002}).

\bibitem[{\citenamefont{Cardy}(1996)}]{Cardy_book96}
\bibinfo{author}{\bibfnamefont{J.}~\bibnamefont{Cardy}},
  \emph{\bibinfo{title}{Scaling and renormalization in statistical physics}}
  (\bibinfo{publisher}{Cambridge University Press},
  \bibinfo{address}{Cambridge}, \bibinfo{year}{1996}).

\bibitem[{\citenamefont{Callan and Coleman}(1977)}]{CallanColeman77}
\bibinfo{author}{\bibfnamefont{C.~G.} \bibnamefont{Callan}} \bibnamefont{and}
  \bibinfo{author}{\bibfnamefont{S.}~\bibnamefont{Coleman}},
  \bibinfo{journal}{Phys. Rev. D} \textbf{\bibinfo{volume}{16}},
  \bibinfo{pages}{1762} (\bibinfo{year}{1977}).

\bibitem[{\citenamefont{Caldeira and Legett}(1983)}]{CaldeiraLeggett83}
\bibinfo{author}{\bibfnamefont{A.~O.} \bibnamefont{Caldeira}} \bibnamefont{and}
  \bibinfo{author}{\bibfnamefont{A.~J.} \bibnamefont{Legett}},
  \bibinfo{journal}{Ann. Phys. (N.Y.)} \textbf{\bibinfo{volume}{149}},
  \bibinfo{pages}{374} (\bibinfo{year}{1983}).

\bibitem[{\citenamefont{Leggett et~al.}(1987)\citenamefont{Leggett,
  Chakravarty, Dorsey, Fisher, Garg, and Zwerger}}]{LCDFGZ87}
\bibinfo{author}{\bibfnamefont{A.~J.} \bibnamefont{Leggett}},
  \bibinfo{author}{\bibfnamefont{S.}~\bibnamefont{Chakravarty}},
  \bibinfo{author}{\bibfnamefont{A.~T.} \bibnamefont{Dorsey}},
  \bibinfo{author}{\bibfnamefont{M.~P.~A.} \bibnamefont{Fisher}},
  \bibinfo{author}{\bibfnamefont{A.}~\bibnamefont{Garg}}, \bibnamefont{and}
  \bibinfo{author}{\bibfnamefont{W.}~\bibnamefont{Zwerger}},
  \bibinfo{journal}{Rev. Mod. Phys.} \textbf{\bibinfo{volume}{59}},
  \bibinfo{pages}{1} (\bibinfo{year}{1987}).

\bibitem[{\citenamefont{Bean and Livingston}(1959)}]{BeanLivingston59}
\bibinfo{author}{\bibfnamefont{C.~P.} \bibnamefont{Bean}} \bibnamefont{and}
  \bibinfo{author}{\bibfnamefont{J.~D.} \bibnamefont{Livingston}},
  \bibinfo{journal}{J. Appl. Phys.} \textbf{\bibinfo{volume}{30}},
  \bibinfo{pages}{S120} (\bibinfo{year}{1959}).

\bibitem[{\citenamefont{Chudnovsky and Gunther}(1988)}]{ChudnovskyGunther88}
\bibinfo{author}{\bibfnamefont{E.~M.} \bibnamefont{Chudnovsky}}
  \bibnamefont{and} \bibinfo{author}{\bibfnamefont{L.}~\bibnamefont{Gunther}},
  \bibinfo{journal}{Phys. Rev. Lett.} \textbf{\bibinfo{volume}{60}},
  \bibinfo{pages}{661} (\bibinfo{year}{1988}).

\bibitem[{\citenamefont{Stauffer}(1976)}]{Stauffer76}
\bibinfo{author}{\bibfnamefont{D.}~\bibnamefont{Stauffer}},
  \bibinfo{journal}{Sol. State. Commun.} \textbf{\bibinfo{volume}{18}},
  \bibinfo{pages}{533} (\bibinfo{year}{1976}).

\bibitem[{\citenamefont{Hoyos and Vojta}()}]{HoyosVojta_unpublished}
\bibinfo{author}{\bibfnamefont{J.~A.} \bibnamefont{Hoyos}} \bibnamefont{and}
  \bibinfo{author}{\bibfnamefont{T.}~\bibnamefont{Vojta}},
  \bibinfo{note}{unpublished}.

\bibitem[{\citenamefont{Senthil and Sachdev}(1996)}]{SenthilSachdev96}
\bibinfo{author}{\bibfnamefont{T.}~\bibnamefont{Senthil}} \bibnamefont{and}
  \bibinfo{author}{\bibfnamefont{S.}~\bibnamefont{Sachdev}},
  \bibinfo{journal}{Phys. Rev. Lett.} \textbf{\bibinfo{volume}{77}},
  \bibinfo{pages}{5292} (\bibinfo{year}{1996}).

\bibitem[{\citenamefont{Dorsey et~al.}(1986)\citenamefont{Dorsey, Fisher, and
  Wartak}}]{DorseyFisherWartak86}
\bibinfo{author}{\bibfnamefont{A.~T.} \bibnamefont{Dorsey}},
  \bibinfo{author}{\bibfnamefont{M.~P.~A.} \bibnamefont{Fisher}},
  \bibnamefont{and} \bibinfo{author}{\bibfnamefont{M.~S.}
  \bibnamefont{Wartak}}, \bibinfo{journal}{Phys. Rev. A}
  \textbf{\bibinfo{volume}{33}}, \bibinfo{pages}{1117} (\bibinfo{year}{1986}).

\bibitem[{\citenamefont{Schehr and Rieger}(2006)}]{SchehrRieger06}
\bibinfo{author}{\bibfnamefont{G.}~\bibnamefont{Schehr}} \bibnamefont{and}
  \bibinfo{author}{\bibfnamefont{H.}~\bibnamefont{Rieger}},
  \bibinfo{journal}{Phys. Rev. Lett.} \textbf{\bibinfo{volume}{96}},
  \bibinfo{pages}{227201} (\bibinfo{year}{2006}).

\bibitem[{\citenamefont{Hoyos and Vojta}(2006)}]{HoyosVojta06}
\bibinfo{author}{\bibfnamefont{J.~A.} \bibnamefont{Hoyos}} \bibnamefont{and}
  \bibinfo{author}{\bibfnamefont{T.}~\bibnamefont{Vojta}},
  \bibinfo{journal}{Phys. Rev. B} \textbf{\bibinfo{volume}{74}},
  \bibinfo{pages}{140401} (\bibinfo{year}{2006}).

\bibitem[{\citenamefont{Dobrosavljevic and
  Miranda}(2005)}]{DobrosavljevicMiranda05}
\bibinfo{author}{\bibfnamefont{V.}~\bibnamefont{Dobrosavljevic}}
  \bibnamefont{and} \bibinfo{author}{\bibfnamefont{E.}~\bibnamefont{Miranda}},
  \bibinfo{journal}{Phys. Rev. Lett.} \textbf{\bibinfo{volume}{94}},
  \bibinfo{pages}{187203} (\bibinfo{year}{2005}).

\end{thebibliography}

\end{document}